# Detection of Scintillation Light of Liquid Xenon with a LAAPD

V.N. Solovov*[1], A. Hitachi[2,3], V. Chepel[3], M.I. Lopes[3], R. Ferreira Marques[3] and A.J.P.L. Policarpo[3]

[1]Department of Physics of the University of Coimbra, 3004-516 Coimbra, Portugal

[2]Kochi Medical School, Nankokushi, Kochi 783-8505, Japan

[3]LIP-Coimbra and Department of Physics of the University of Coimbra, 3004-516 Coimbra, Portugal

**Abstract**

First observation of liquid xenon scintillation due to $\alpha$-particles and $\gamma$-rays with a large area avalanche photodiode immersed in the liquid is reported. An energy resolution of 10% (FWHM) and a coincidence time resolution of less then 1 ns (FWHM) were measured with 5.5 MeV $\alpha$-particles and 511 keV $\gamma$-rays, respectively.

The quantum efficiency of the photodiode for xenon scintillation light ($\lambda$=178 nm) is estimated to be ~100%.

---

* Correponding author: solovov@lipc.fis.uc.pt

# 1. Introduction

Liquid xenon is a promising medium for detectors of ionising radiation. Large atomic number, relatively high density, high light output and fast decay time make it an excellent scintillator suitable for a number of applications such as particle physics, astrophysics and nuclear medicine. Its scintillation spectrum lies in the short wavelength region centred at $\lambda=178$ nm [1]. Besides being sensitive to the VUV light, the photodetectors must perform under rather severe conditions, namely, low temperature (about -100ºC) and external pressure variations typically in the range from $10^{-7}$ mbar to 5 bar. In addition, they have to be compatible with high purity liquid xenon environment.

Several groups used successfully some types of photomultiplier tubes in liquid xenon. Recently developed large area avalanche photodiodes (LAAPD) have higher quantum efficiency, are compact and do not need voltage dividers. Moreover, low inherent radioactivity and small mass can be an additional benefit for some applications (search for Weakly Interactive Massive Particles is an example [2]).

It was shown that good energy and time resolution can be obtained with a LAAPD coupled to crystal scintillators [3,4]. A LAAPD was also operated in a gas scintillation proportional counter filled with xenon gas [5]. In our previous work [6], we have demonstrated that LAAPDs can be operated at temperatures down to -100 °C without degradation of their performance. Moreover, the dark current decreases with decreasing temperature by more than four orders of magnitude with respect to room temperature.

In the present paper we report on the first observation of liquid xenon scintillation due to $\alpha$-particles and $\gamma$-rays with a large area avalanche photodiode.

## 2. Experimental Set-up

We used a windowless LAAPD from Advanced Photonix, Inc. [7] with a sensitive area of 0.2 cm$^2$. These LAAPDs, manufactured using beveled-edge technology, are sensitive in the wavelength region from 130 to 1100 nm. They have high quantum efficiency, low dark current and relatively small capacitance (~25 pF at a gain of 200 [8]).

The LAAPD was placed in a small stainless steel chamber (Fig.1) connected to the gas purification system described elsewhere [9]. The chamber was heated to 70°C (maximum storage temperature for the LAAPD [8]) and pumped down to $3 \times 10^{-7}$ mbar. Then, purified xenon was circulated through the chamber during several hours to remove residual impurities. The chamber was cooled down to -100 °C, approximately, in a bath of alcohol mixed with liquid nitrogen and filled with liquid xenon until the photodiode was completely immersed in the liquid. The temperature during the measurements was stabilized at -102±1°C being monitored by two platinum thermoresistors glued to the middle and to the bottom of the chamber. The photodiode was operated at bias voltage between 400 and 1700 V corresponding, at –102°C, to the gain range from 1 to 1200.

### A. *Energy resolution measurements*

An $^{241}$Am α-source deposited on one side of a stainless steel disk was placed at a distance of 15 mm from the photodiode (Fig.1). The active area of the source was covered with a mask made from stainless steel foil with a hole of 1.5 mm diameter to restrict the solid angle variation. The LAAPD was connected to a low noise charge sensitive preamplifier (Canberra 2001A). The fact that the dark current at -100 °C is less than 1 pA allowed to connect it as an ordinary solid state detector, i.e. with the bias voltage supplied through the highly resistive RC-filter incorporated in the preamplifier. The signal was amplified with a shaping amplifier (Canberra 2021) and fed into a multichannel analyzer. All measurements were made with the optimal shaping constant of 1 μs.

Calibration of the whole spectrometric channel was made with a precision pulser connected to the test input of the preamplifier.

*B. Time resolution measurements*

Time resolution measurements were carried out with a $^{22}$Na source placed between the experimental chamber and a barium fluoride crystal coupled to a Philips XP2020Q photomultiplier. The bias voltage was applied to the LAAPD through an external filter, as shown in Fig.2. The output signal from the LAAPD was fed to a fast voltage preamplifier [10] with 50 Ω input impedance, 3.5 ns risetime and noise of 10 μV r.m.s. referred to the input and split between a fast amplifier (Phillips Scientific 777) and a shaping amplifier (Canberra 2021, with $\tau_i=\tau_d=1\mu s$) for measuring the amplitude. The reference signal was generated by the BaF$_2$ crystal. Constant fraction discriminators (Phillips Scientific 715) were used in both timing channels. The output signals of these two discriminators were used as START and STOP for a time-to-amplitude converter (TAC) ORTEC 566. The amplitudes of the signals at the outputs of the TAC and shaping amplifier were both digitized with a peak ADC (LeCroy 2259B), which was triggered with a gate produced by a coincidence of the START and STOP signals within a time window of 100 ns. The information was stored in a computer for subsequent off-line analysis. This set-up allowed us to measure the amplitude dependence of the time resolution.

**3. Results and Discussion**

*A. Energy resolution and quantum efficiency*

The dependence of the gain on the bias voltage was measured by recording the amplitude of the charge signal from the LAAPD due to the scintillation produced in liquid xenon by α-particles as a function of the bias voltage (Fig. 3). The number of primary electron-hole pairs per α-particle, $N_0$, was determined from measurements carried under applied voltages for which the amplitude of the LAAPD signal was constant and therefore the gain could be assumed unitary. The value of $N_0$ was obtained comparing the amplitude of the signal due to α-particles under this condition with that of a calibration signal produced by injection of a known charge into the input of the preamplifier. A value of $N_0$ equal to 2400±100 was found. The gain was obtained by dividing the output charge of the LAAPD by $eN_0$ ($e$ is the elementary charge).

Pulse height spectra of liquid xenon scintillation due to 5.5 MeV α-particles were acquired for different values of the LAAPD gain. A typical amplitude spectrum obtained with a gain of 120 is shown in Fig.4. In a different geometry (the source at a distance of about 5 mm from the LAAPD window), the 60 keV γ-rays emitted by the $^{241}$Am source could also be detected as shown in Fig.5. The energy resolution is poor due to the large variation of solid angle.

The dependence of the energy resolution obtained with α-particles on the LAAPD gain is presented in Fig.6.

The energy resolution of an APD-scintillator system can be written as [11]:

$$\frac{\Delta E}{E} = 2.355\sqrt{\left(\frac{N_e}{N_0 M}\right)^2 + \frac{F-1}{N_0} + \delta^2} , \qquad (1)$$

where $N_e$ is the number of noise electrons referred to the preamplifier input, M is the APD mean gain and F is the excess noise factor which takes into account the fluctuations inherent to the multiplication process. The excess noise factor can be estimated [12] with

$$F \approx kM + (2 - 1/M)(1-k), \qquad (2)$$

where k is a weighted average ratio of the hole ionization rate to that for electrons, which for a beveled-edge LAAPD at normal conditions is approximately equal to 0.0017 [4]. In eq. (1), the first term is the electronic noise contribution, the second term takes into account the statistical fluctuations of the gain and the third one includes all the fluctuations associated with the scintillation, light collection and photon-to-electron conversion processes.

The value of $N_e$ was determined experimentally with a test pulse fed through a known capacitance to the input of the preamplifier. It varies with the gain due to the fact that the LAAPD capacitance depends on the bias voltage. For M>3, however, it is practically constant and equal to 240 electrons, r.m.s. The shot noise due to dark current of the LAAPD is very small at -100 °C (few electrons), for an integration time constant of ~1µs [6].

As one can see in Fig.6, for gains lower than 10 the resolution is dominated by the preamplifier noise. For gains larger than 200 the resolution becomes worse as the excess noise factor increases with the increasing gain. For M between 10 and 200 the resolution is practically constant, being about 10%.

Eq. (1) was fit to the experimental data with $N_e$, k and δ as free parameters. The best fit was obtained with $N_e$=255±16, k=0.0029±0.0003 and δ=0.0370±0.0006. The value for $N_e$ is in good agreement with the measured one. The value found for k is significantly higher then that usually referred in the literature [4] (k=0.0017) for this type of devices at room temperature. As for the δ value, according to our estimate, the solid angle variation contributes to δ with approximately 0.01. The photoelectron statistics, calculated under the assumption that a photon produces no more than one photoelectron, contributes an additional 0.02 $(1/\sqrt{N_0})$. The missing fluctuations of about 0.029 (in order to obtain 0.037) may arise due to the fact that it is energetically possible for a VUV photon with the energy of 7.1 eV to produce two or more electron-hole pairs. In this case, the above estimate of the fluctuations in the number of photoelectrons is not valid. Eventual non-uniformity of VUV light reflection from the α-source surface may also contribute.

From the above results one can estimate the quantum efficiency of the LAAPD, Q, understood as the average number of primary electron-hole pairs produced by an incident photon. If we neglect reflections of VUV from the walls and the α-source surface, the number of electron-hole pairs due to an α-particle is given by

$$N_0 = \frac{E_\alpha}{W_s} \frac{\Omega}{4\pi} Q \qquad (4)$$

where $E_\alpha$ is the energy of the α-particle, $W_s$ the average energy required per one scintillation photon in liquid xenon and $\Omega$ the solid angle.

Taking $W_s$=16.3eV [13], $E_\alpha$=5.5×10$^6$ eV, $4\pi/\Omega$≈150 and $N_0$=2400 one gets Q ≈ 100%.

*B. Time resolution*

For γ-photon irradiation in the geometry of Fig.2, the solid angle and thereby the LAAPD signal amplitude depend considerably on the position where the scintillation is produced in liquid xenon. Fig. 7 shows the time interval distributions for the signals above the threshold of 1,500 primary electron-hole pairs $N_0$ (a), and for the signals in the range of $N_0$ from 6,000 to 8,000 (b). For further analysis, the events were grouped according to their amplitude and the width of the time interval distribution was found for each amplitude group. This procedure allowed measuring the time resolution as a function of $N_0$ in the range from 1,000 to 10,000 primary electron-hole pairs. Fig. 8 shows the results obtained for gains of 280, 605 and 1208. In all cases, the discriminator threshold referred to the input of the preamplifier was set to 50 μV, i.e. five standard deviations above the preamplifier noise. For lower values of the threshold, the rate of random coincidences was too high.

For a given gain, the time resolution improves with the increase of the number of photons reaching the photodiode. For fixed $N_0$, it is better for higher gains. The best resolution, 0.9 ns fwhm, was achieved with the gain of 605 at $N_0$≈7800.

The data obtained for gain M=280 is in good agreement with the results for LSO and YAP scintillator crystals coupled to a LAAPD similar to the one we used [3] (see Table 1). These crystals have decay time constants of 40 ns and 26.7 ns respectively, being close to that of liquid xenon (45 ns [14]). Hence, we can conclude that the LAAPD maintains its good timing properties at low temperature.

The signals due to direct interaction of 511 keV γ-rays with the LAAPD were also observed. These events are, apparently, responsible for a peak in the coincidence time distribution preceding that due to xenon scintillation by about 4 ns (Fig. 9).

**4. Conclusions**

A LAAPD was used for the detection of scintillation photons in liquid xenon. Immersed into the liquid, it has proven to be operational at T = -100 °C.

An energy resolution of 10% (FWHM) was obtained with 5.5 MeV α-particles.

The coincidence time resolution, measured with 511 keV γ-rays, is similar to that reported for LSO and YAP scintillation crystals. The best value, obtained for the LAAPD gain of 605, is 0.9 ns (fwhm).

The estimated quantum efficiency for liquid xenon scintillation photons is about 100 %

**Acknowledgements**

This work was financed by the project CERN/P/FIS/1594/1999 from the Fundação para a Ciência e Tecnologia, Portugal. One of the authors was supported by a fellowship PRAXIS XXI/BD/3892/96 from the same organisation.

Akira Hitachi received a fellowship from Fundação Oriente, Portugal.

| $N_0$ | Liquid xenon (M=280) | Crystals (M=250) |
|---|---|---|
| 6,300 | 1.6 | 1.71 (YAP) |
| 11,000 | 1.1 | 1.26 (LSO) |

Table 1: Comparison of time resolution obtained in the present work with values measured with scintillation crystals [3], for the same $N_o$.

**Figure Captions**

Figure 1: Set-up for energy resolution measurements: PA – charge sensitive preamplifier, Amp – spectroscopy amplifier, MCA – multichannel analyser.

Figure 2: Set-up for time resolution measurements: PA – preamplifier, FO – linear fan-out, SA – shaping amplifier, FA – fast amplifier, CFD – constant fraction discriminator, C – coincidence unit, DL – delay line, TAC – time-to-amplitude converter, GG – gate generator.

Figure 3: LAAPD gain as a function of bias voltage (T = -100ºC).

Figure 4: Typical pulse height spectrum of the scintillation due to 5.5 MeV α-particles. LAAPD gain is 120.

Figure 5: The pulse height spectrum obtained with the $^{241}$Am source placed at 5mm from the photodiode. A peak due to 60 keV γ-rays can be distinguished. The LAAPD gain is 150. The amplifier gain is 3 times higher than for the spectrum shown in Fig. 4.

Figure 6: Energy resolution measured with α-particles as a function of gain. The squares are experimental points and the line is the best fit of eq. (1) to the experimental data.

Figure 7: Time interval distributions, obtained with the LAAPD gain M=605 for $N_0$ above the threshold of 1,500 (a) and for $N_0$ in the range from 6,000 to 8,000 (b).

Figure 8: Time resolution as a function of the number of primary electron-hole pairs ($N_0$) for three values of the LAAPD gain.

Figure 9: Time interval distribution including events involving direct interaction of γ-rays with the LAAPD (left peak) and those due to the scintillation produced in liquid xenon.

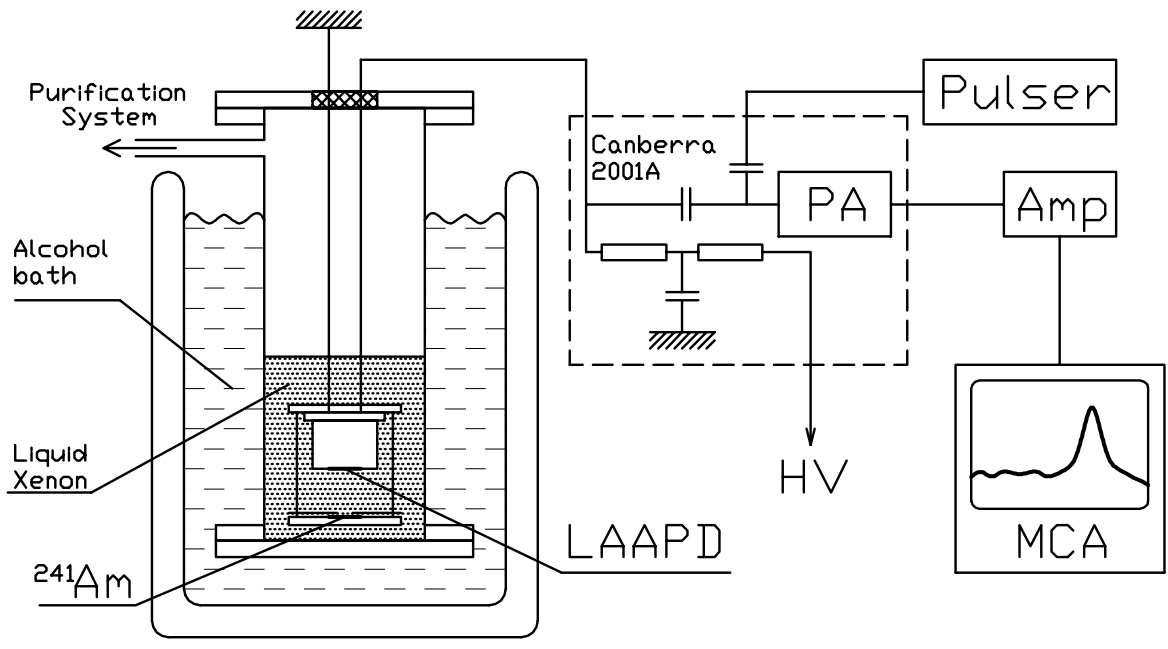

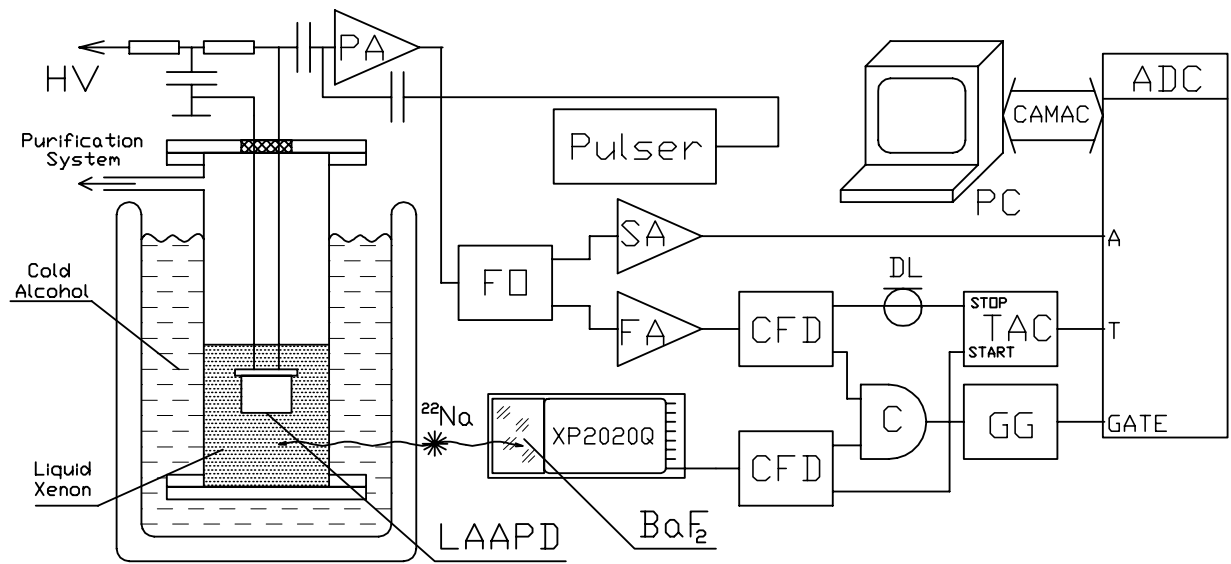

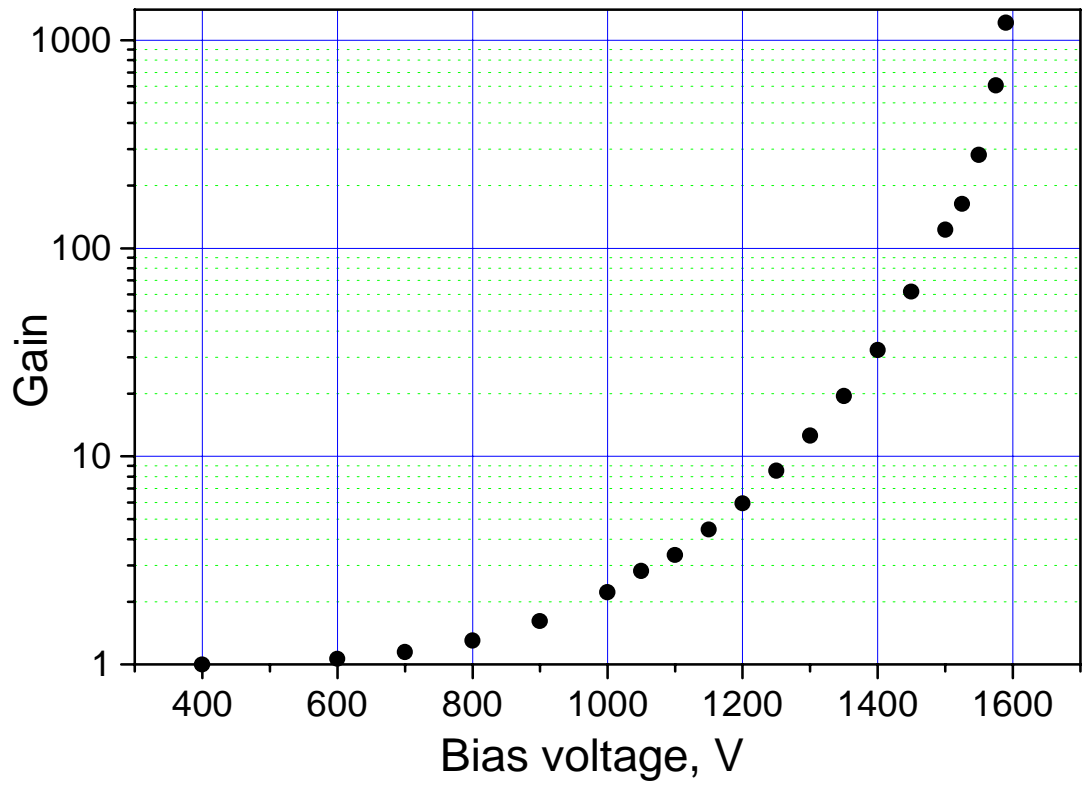

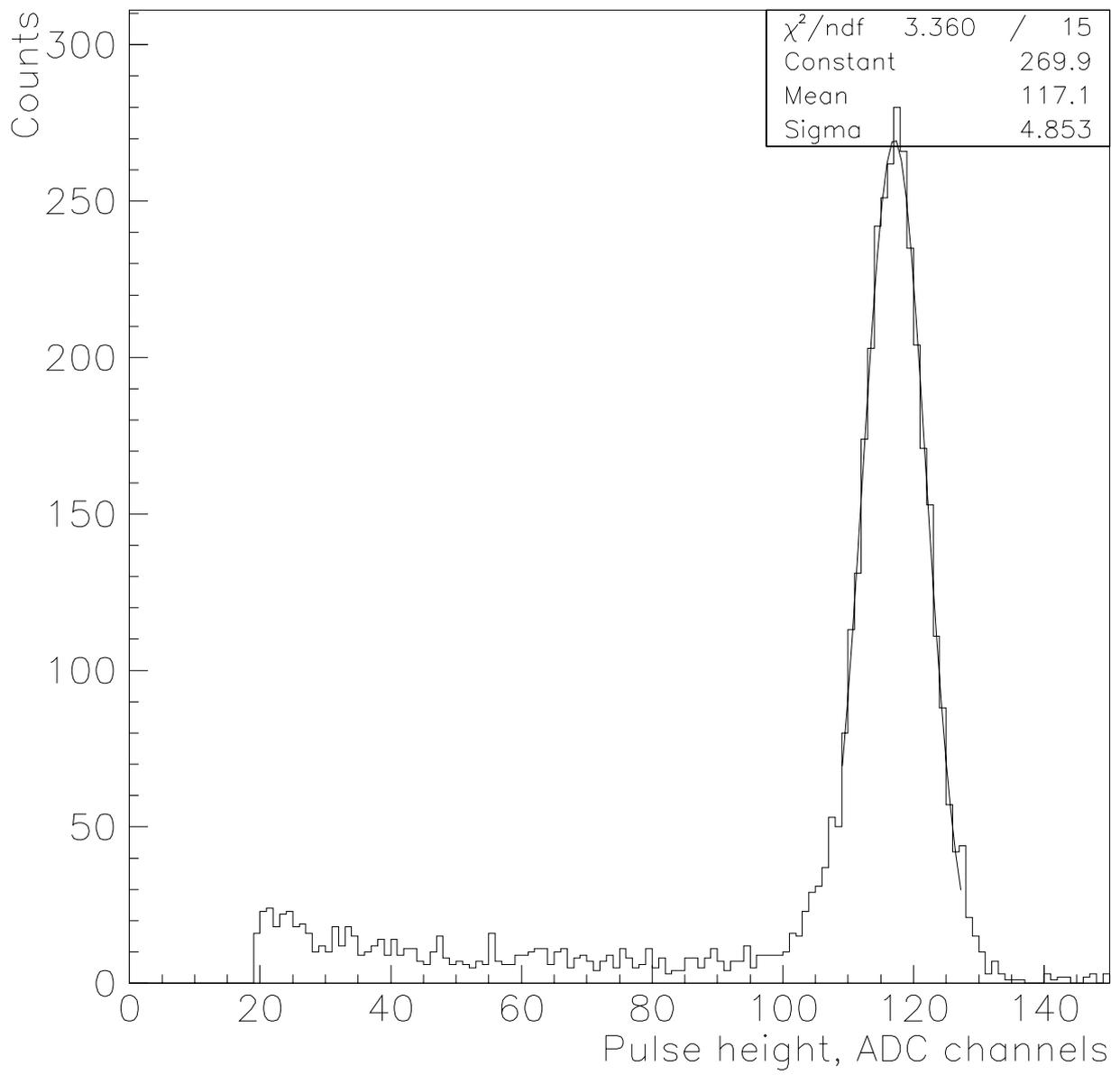

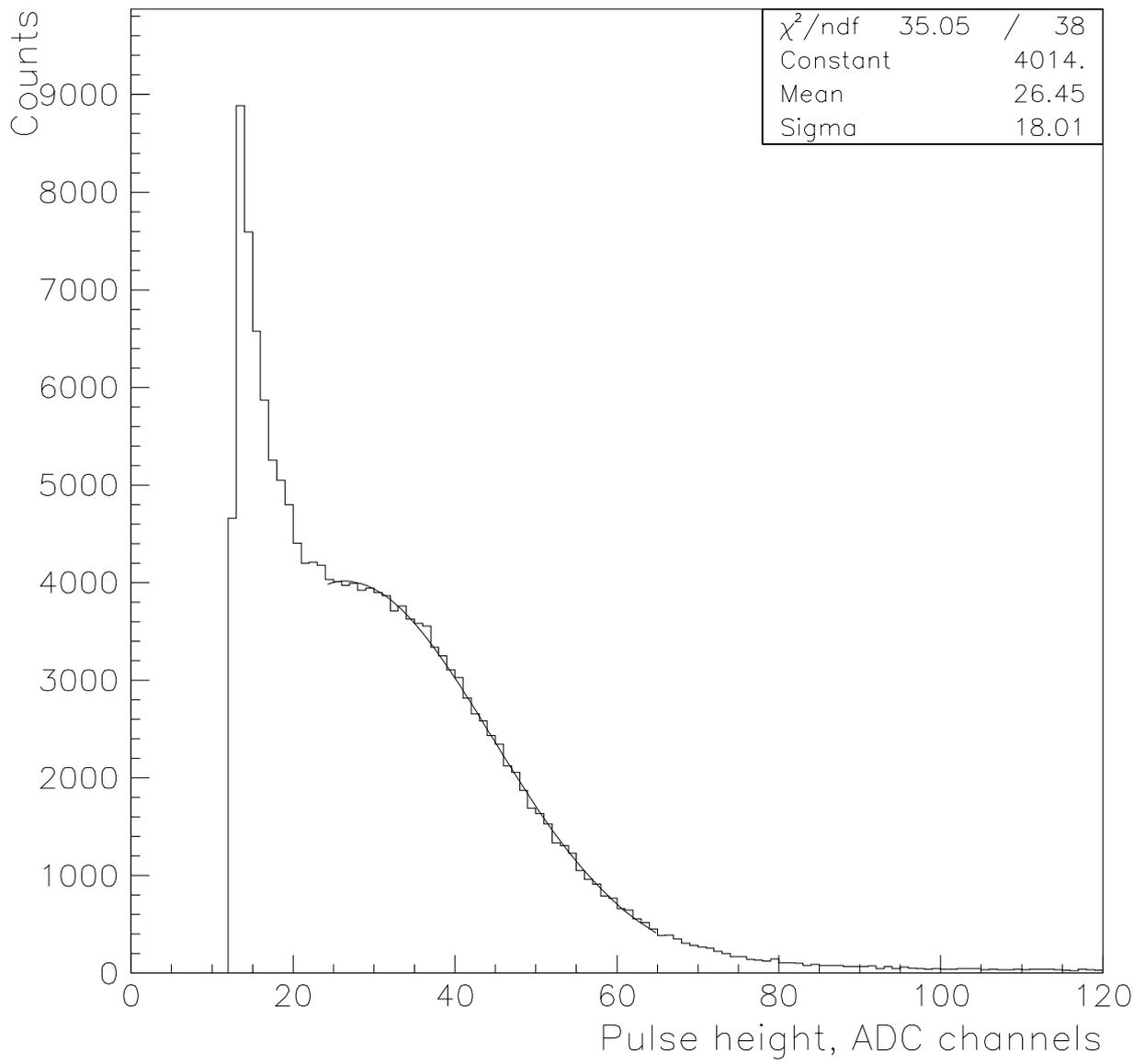

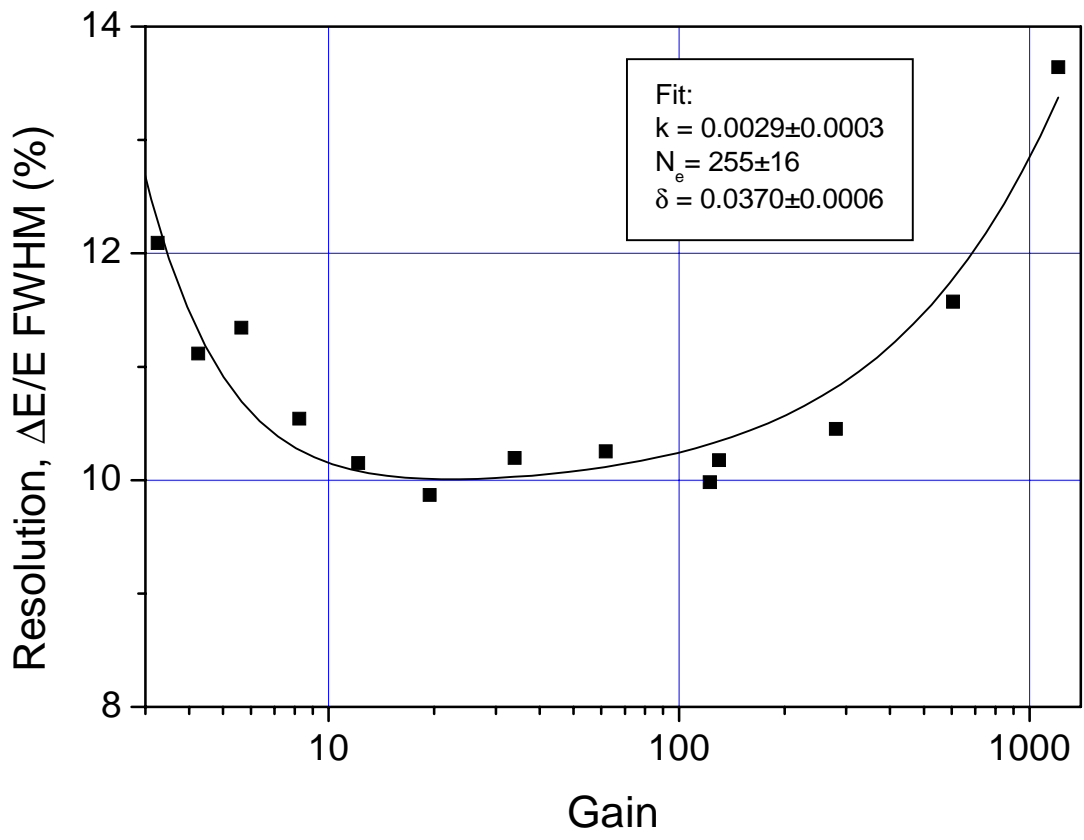

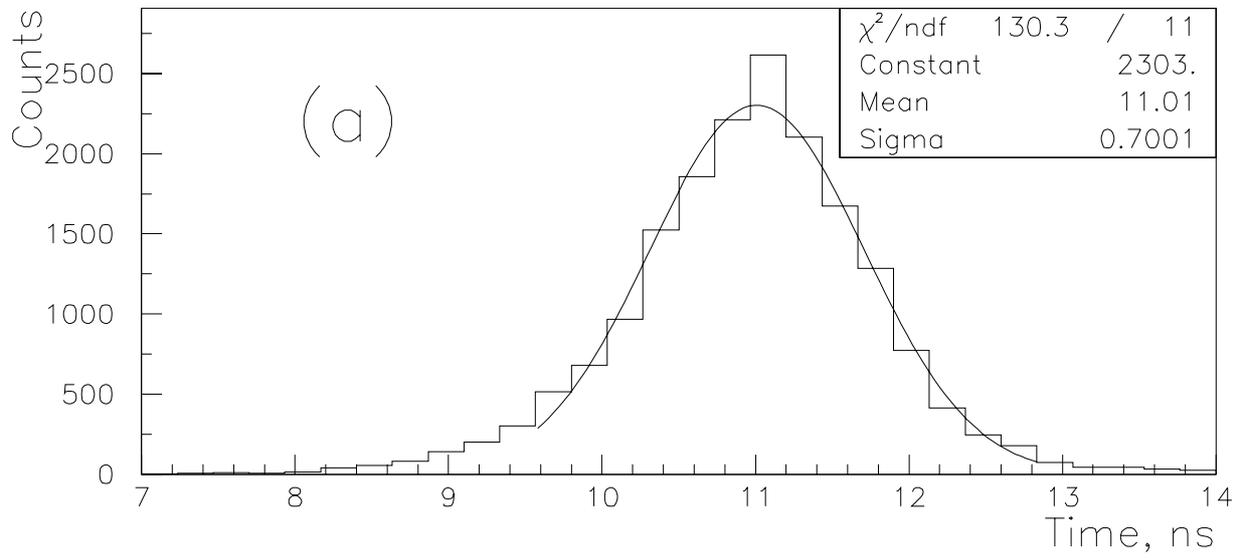

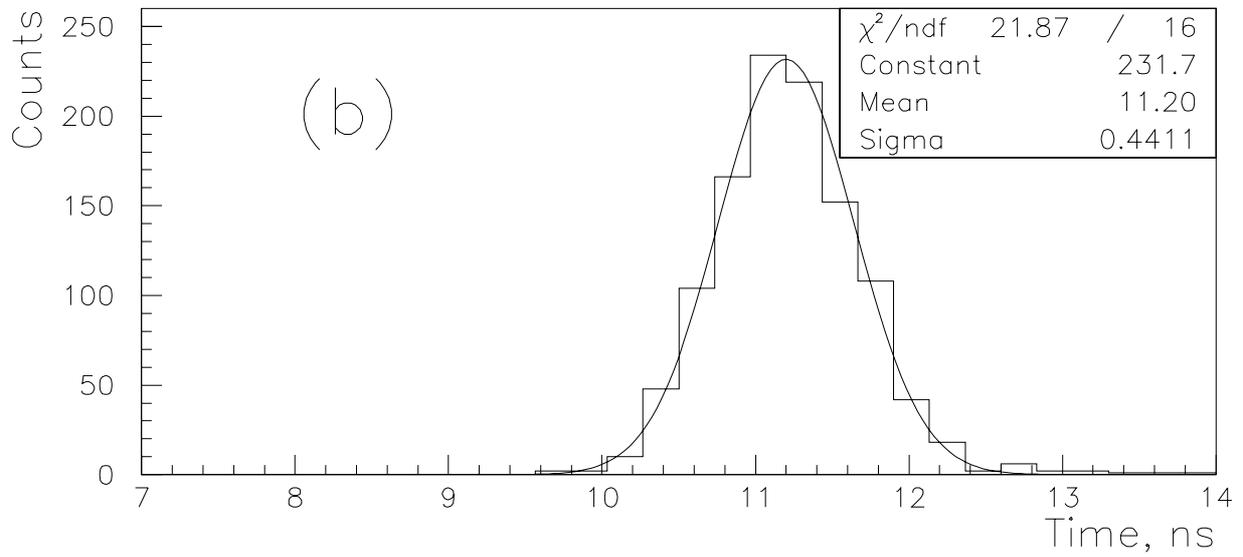

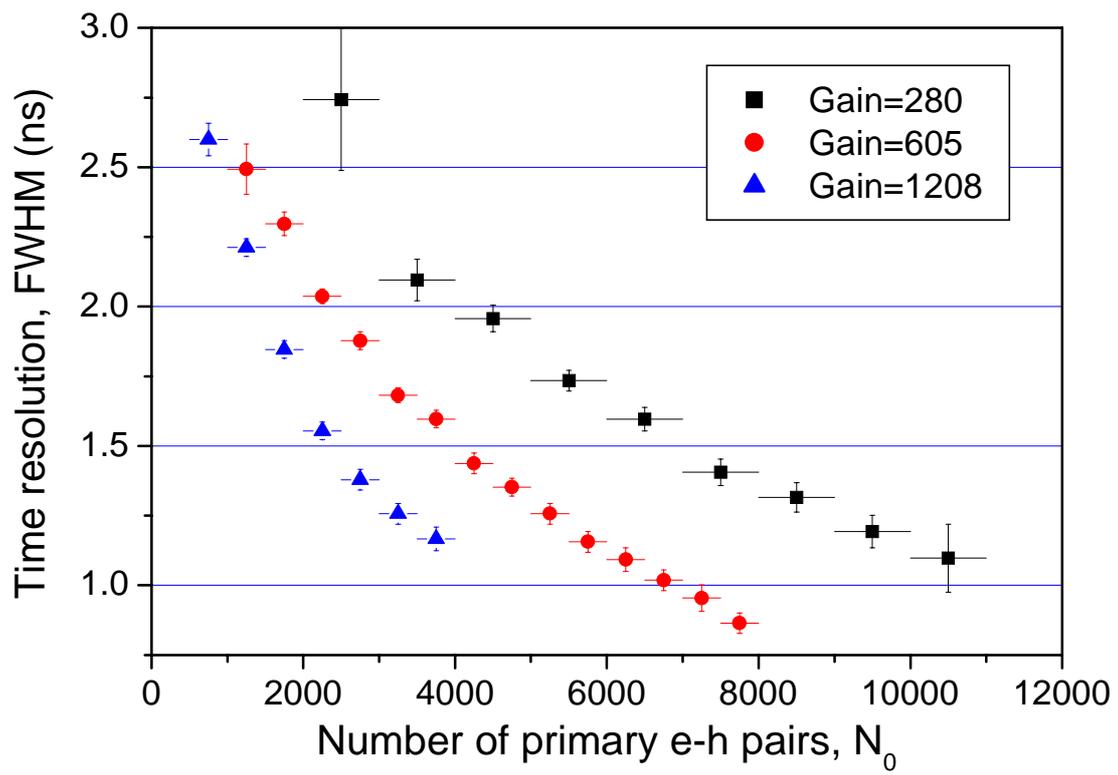

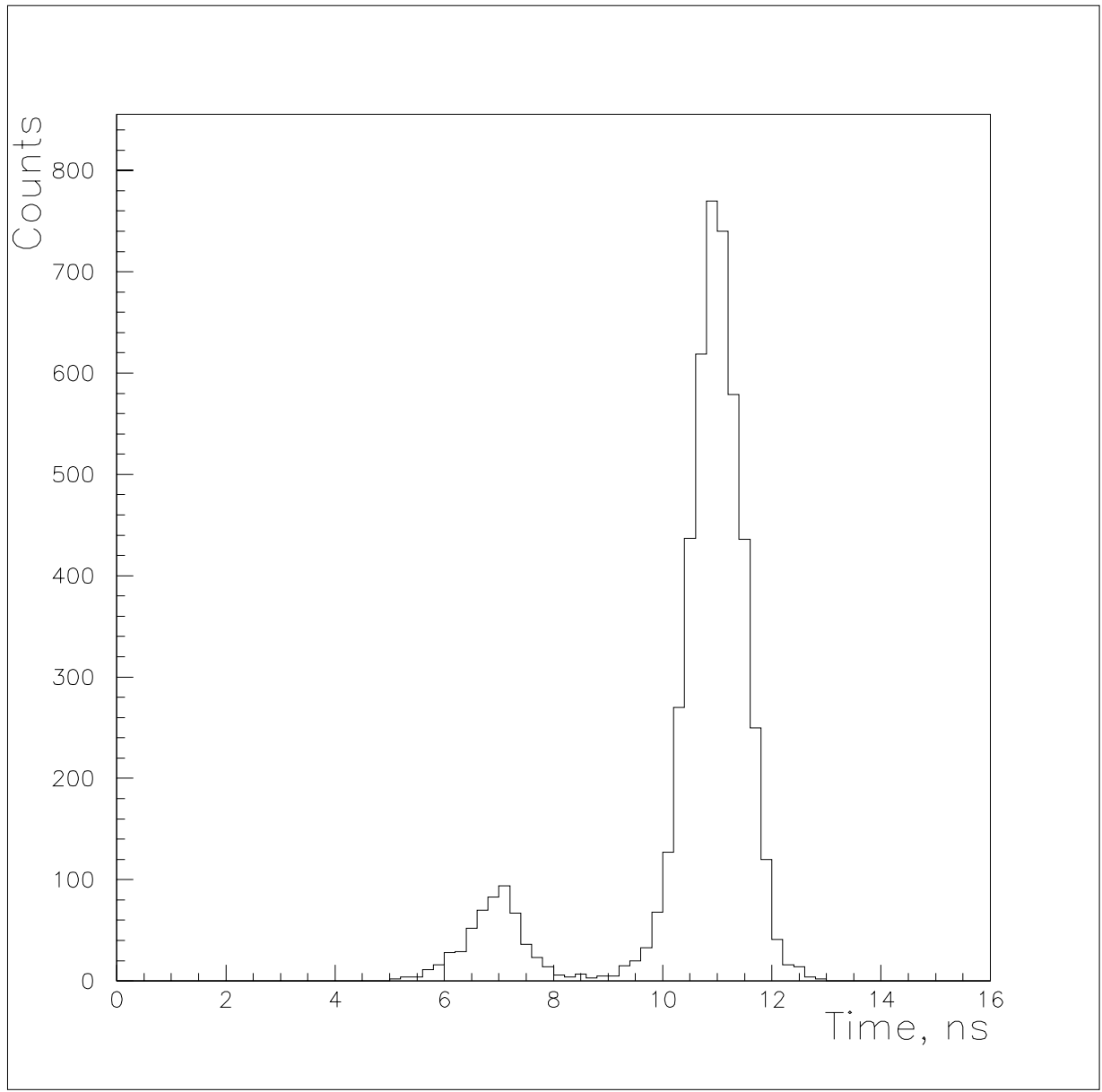